\pdfoutput=1
\documentclass[aps,prl,reprint,groupedaddress]{revtex4-1}
\usepackage{graphicx}  
\usepackage{dcolumn}   
\usepackage{bm}        
\usepackage{amssymb}   
\usepackage{amsmath}   
\usepackage{graphics,color}
\usepackage{hyperref}
\hypersetup{
    unicode=false,          
    pdftoolbar=true,        
    pdfmenubar=true,        
    pdffitwindow=false,     
    pdfstartview={FitH},    
    pdftitle={My title},    
    pdfauthor={Author},     
    pdfsubject={Subject},   
    pdfcreator={Creator},   
    pdfproducer={Producer}, 
    pdfkeywords={keyword1} {key2} {key3}, 
    pdfnewwindow=true,      
    colorlinks=true,       
    urlcolor=black,
    linkcolor=black,          
    citecolor=blue,        
    filecolor=black,      
    urlcolor=black           
}
\usepackage[greek,english]{babel}
\newcommand{\gm}{\greektext m\latintext}

\begin{document}


\title{Lifetimes of Confined Acoustic Phonons in Ultra-Thin Silicon Membranes}


\author{J. Cuffe$^1$}
\email[]{john.cuffe@icn.cat}
\author{E. Ch\'{a}vez$^{1,2}$}
\author{P-O. Chapuis$^1$}

\author{F. Alzina$^1$}
\author{C. M. Sotomayor Torres$^{1,2,3}$}
\email[]{clivia.sotomayor@icn.cat}

\affiliation{$^1$Catalan Institute of Nanotechnology (ICN2), Campus UAB, 08193 Bellaterra (Barcelona), Spain}
\affiliation{$^2$Department of Physics, UAB, 08193 Bellaterra (Barcelona), Spain}
\affiliation{$^3$Institució Catalana de Recerca i Estudis Avançats (ICREA), 08010 Barcelona, Spain}

\author{O. Ristow}
\author{M. Hettich}
\author{T. Dekorsy}
\affiliation{Department of Phyics and Center of Applied Photonics, Universitaet Konstanz, D-78457 Konstanz, Germany}

\author{A. Shchepetov}
\author{M. Prunnila}
\author{J. Ahopelto}
\affiliation{VTT Technical Research Centre of Finland, PO Box 1000, 02044 VTT, Espoo, Finland}

\date{\today}

\begin{abstract}
We study the relaxation of coherent acoustic phonon modes with frequencies up to 500 GHz in ultra-thin free-standing silicon membranes. Using an ultrafast pump-probe technique of asynchronous optical sampling, we observe that the decay time of the first-order dilatational mode decreases significantly from $\sim$ 4.7 ns to 5 ps with decreasing membrane thickness from $\sim$ 194 to 8 nm. The experimental results are compared with theories considering both intrinsic phonon-phonon interactions and extrinsic surface roughness scattering including a wavelength-dependent specularity. Our results provide insight to understand some of the limits of nanomechanical resonators and thermal transport in nanostructures.
\end{abstract}



\maketitle

Mechanical and acoustic properties in the nanoscale are receiving increasing attention as they are key properties affecting the limits of ultrasensitive detectors of force \cite{Mamin2001}, mass \cite{Chaste2012,Jensen2008}, charge \cite{Lassagne2009a,Steele2009} and spin \cite{Rugar2004a}, influencing platforms for biosensing \cite{Arlett2011} and the investigation of quantum behaviour in extended objects \cite{O'Connell2010}. In particular, phonon lifetimes influence the achievable mechanical quality (\textit{Q}) –-factors in nanomechanical resonators, which often limit device performance \cite{Unterreithmeier2010}. Moreover, they are necessary input parameters for accurate calculations of nanoscale thermal transport, with high-impact applications such as heat management in nanoelectronics \cite{Pop2010} and the engineering of novel thermoelectric materials \cite{Tang2010}. Despite their importance, phonon lifetimes are perhaps the least well known of all phonon properties due to the challenges associated with their quantitative determination and theoretical modelling. Even though silicon is the most important material for nanoelectronics, MEMS and NEMS, there are few experimental reports of direct measurements of phonon lifetimes in the gigahertz to terahertz range \cite{Daly2009} and for all materials open questions remain about the relative contributions of intrinsic and extrinsic scattering processes at high frequencies in both bulk and nanoscale structures \cite{Ayrinhac2011a,Baldi2010a,Duquesne2003,Liu2007,Unterreithmeier2010}. Recent experimental investigations of phonons in superlattice cavities with frequencies of around 1 THz have suggested that lifetimes of high-frequency phonons could be limited by an average interface roughness of just 0.06 nm \cite{Rozas2009}. On the other hand, phonon wavepackets experiments in bulk silicon with frequencies up to approximately 100 GHz were analysed with a simplified Akhiezer relaxation damping model \cite{Akhiezer1939,Daly2009} of intrinsic scattering, using an average lifetime of high-frequency thermal phonons of 17 ps. Other intrinsic damping models include clamping losses \cite{Wilson-Rae2011}, thermoelastic dissipation \cite{Lifshitz2000a} and three-phonon interactions \cite{AlShaikhi2007}, which predict a different behaviour depending on the frequency and temperature regimes. In this context, generation and detection of coherent acoustic phonons at high frequencies in different materials and nanostructures is an ideal method to obtain quantitative information on phonon lifetimes and compare with the main theoretical models.

Here we use free-standing single-crystalline silicon membranes fabricated by back-etching (100)-oriented silicon-on-insulator (SOI) wafers to study the decay of coherent phonons. These membranes are model systems for such studies, as they can be fabricated with precisely controlled dimensions and physical parameters, facilitating comparison with theoretical models since the analysis is free from interplay with a substrate. This type of membrane was used previously to observe confined acoustic phonons \cite{Torres2004} and study their dispersion relation \cite{Cuffe2012} using inelastic light scattering. We use the ultrafast pump-probe technique of high-speed asynchronous optical sampling (ASOPS) to generate and detect coherent acoustic phonons \cite{Bartels2007}, without the use of any transducing metallic layer. We perform measurements over a large range of thickness values from 7.7 $\pm$ 0.5 to 194 $\pm$ 1 nm, allowing us to investigate the trend in phonon lifetime with frequency up to $\sim$ 500 GHz and compare with predictive models. We compare the experimental results with theories involving intrinsic phonon-phonon interactions and extrinsic surface roughness scattering with a wavelength-dependent specularity parameter.

The ASOPS experiments were performed at room temperature in reflection geometry. The spot size on the membranes was about 1.75 \gm m in diameter and the wavelengths used for pump and probe beams were 780 and 810 nm, respectively. Due to the large optical penetration depth of approximately 8 \gm m, the pump pulse causes a symmetric strain in the membrane via thermal expansion and the hydrostatic deformation potential \cite{Hudert2009,Wright1995}. As a consequence, the first-order dilatational mode at $q_{\parallel}=0$, $D^{1}_{q_{\parallel}=0}$, is excited in the illuminated region, which oscillates at a frequency of $\omega = \pi v_L / d$ , where $v_L$ = 8433 m s$•^{-1}$ is the longitudinal velocity and d is the thickness of the membrane. This mode is identified in the dispersion relation in Fig.\ 1(a). The dilatational oscillation changes the optical cavity thickness of the membrane, which in turn modulates the reflectivity according to well-known Fabry-Perot effects. Even though the change in membrane thickness is of the order of 1 pm and below (Fig.\ 1(b)), corresponding to a small change in reflectivity of about one part in 10$^{-5}$, the ASOPS system is sensitive enough to detect these small changes in reflectivity. A change in reflectivity is also caused by the photoelastic effect; however the change of the optical cavity thickness is the dominant contribution, owing to the small photoelastic constants of silicon. The subsequent dynamics of the confined phonons are then observed by recording the light modulation induced by phonon-photon coupling in a one-dimensional photo-acoustic cavity.

\begin{figure}
 \includegraphics{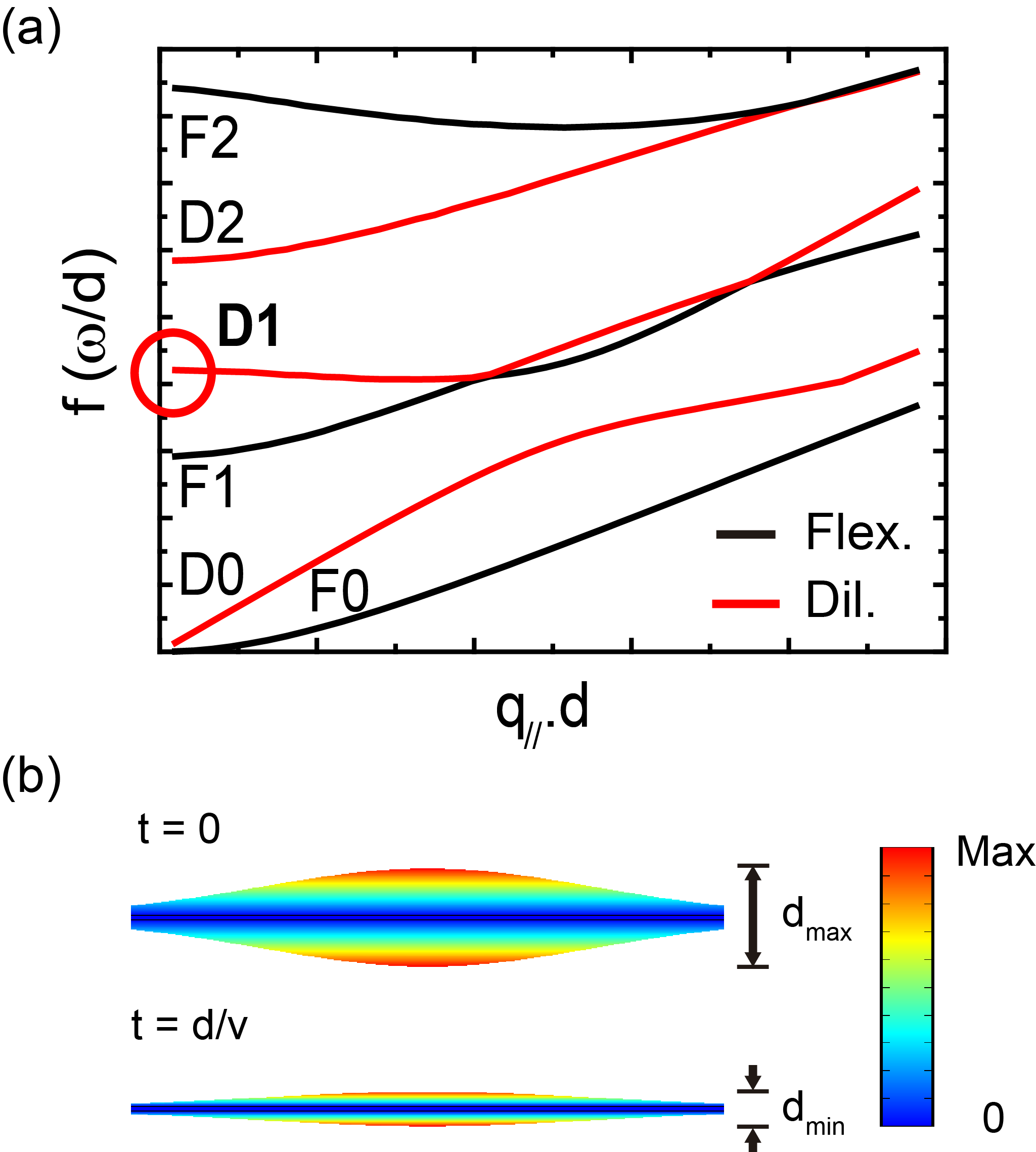}%
 \caption{\label{Fig1}(a) Dispersion relation of a free-standing silicon membrane, showing flexural () and dilatational () modes. The mode primarily excited by the pump pulse is the first-order dilatational mode at zero parallel wavevector, $D^{1}_{q_{\parallel}=0}$. (b) Schematic diagram of the displacement field of the excited mode calculated by finite element simulations. The change in thickness $d$ due to the oscillation of the $D^{1}_{q_{\parallel}=0}$ mode is of the order of 1 pm and below, with a corresponding change in reflectivity of the order of $10^{-5}$. }
 \end{figure}

Figure 2 shows typical time traces of the reflectivity signal from silicon membranes with thickness values of 30 and 100 nm after excitation. At short times, the fast electronic response of the membrane is observed. The electronic contribution can be modelled by a bi-exponential decay and subtracted to reveal the acoustic modes \cite{Bruchhausen2011}, shown in the inset. The decay of the excited coherent phonons is then modelled as a damped harmonic oscillator of the form $\dfrac{\Delta R}{R}\left(t\right) = A \sin (\omega t) \exp(-t/\tau)$ to extract a single phenomenological decay time $\tau$. The obtained lifetimes are plotted in Fig.\ 3 and compared to reported values for bulk silicon \cite{Daly2009} and previous results for a 222 nm silicon membrane \cite{Bruchhausen2011}. The frequencies are those of the $D^{1}_{q_{\parallel}=0}$ mode, which increase with decreasing membrane thickness. It is observed that the lifetimes of coherent phonons in thin silicon membranes decrease dramatically with increasing frequency (decreasing thickness) and do not exhibit a simple behavior as a function of frequency.

 \begin{figure}
 \includegraphics[width=0.45\textwidth]{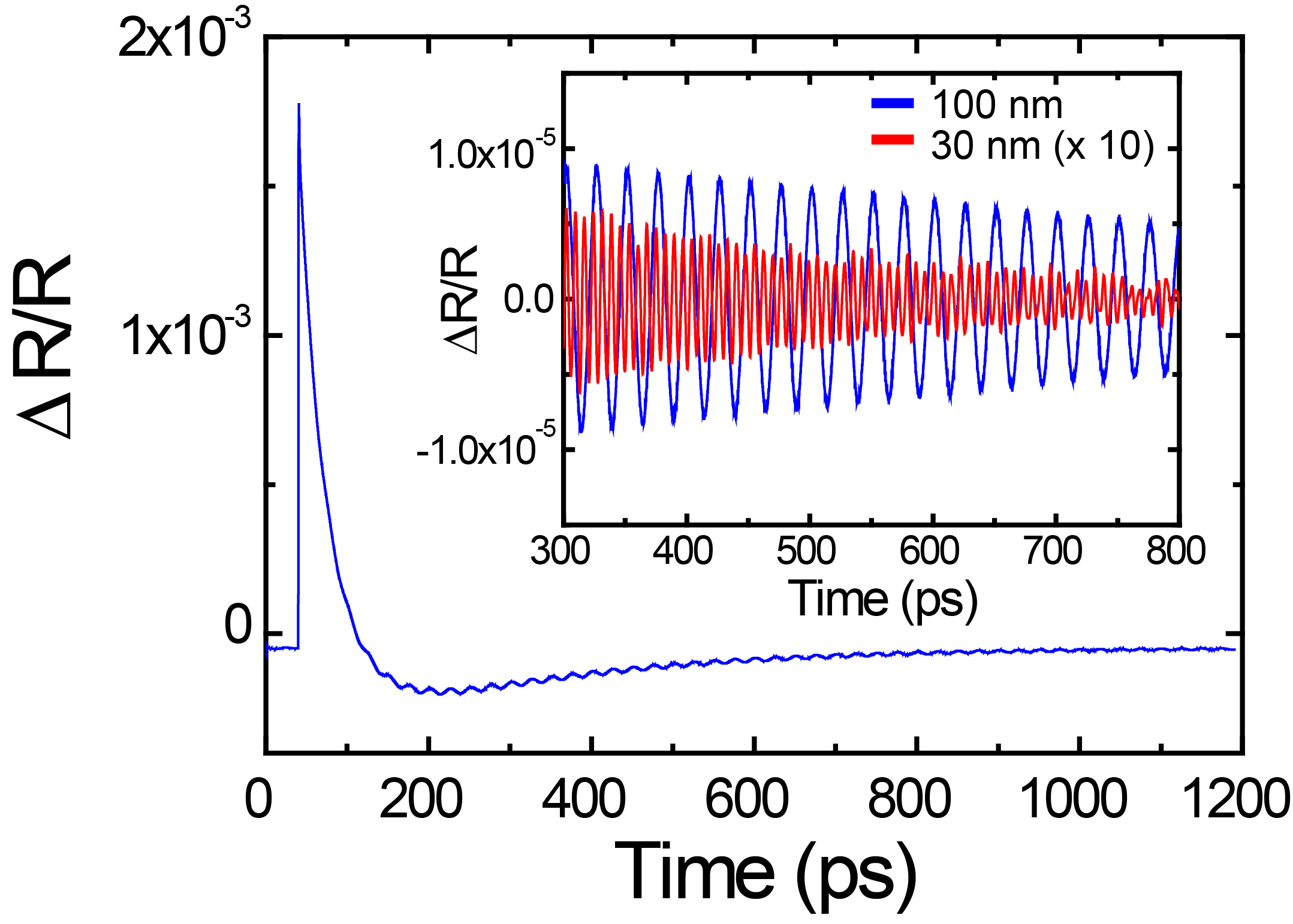}%
 \caption{\label{Fig2}Fractional change in reflectivity as a function of time $(\Delta R / R (t))$ for the 100 nm silicon membrane. The sharp initial change in reflectivity is due to the electronic response of the membrane. The subsequent weaker oscillations are due to the excited acoustic modes. Inset: Close-up of $\Delta R / R (t)$ due to the acoustic modes after subtraction of the electronic response for membranes with 100 and 30 nm thickness shown by a blue and a red line, respectively. The sinusoidal decay of the reflectivity due to the first-order dilatational mode is clearly observed as a function of time, with a faster decay observed for the thinner membrane. The time trace of the 30 nm membrane has been magnified by a factor of 10 for clarity. }
 \end{figure}
 
 \begin{figure}
 \includegraphics[width=0.45\textwidth]{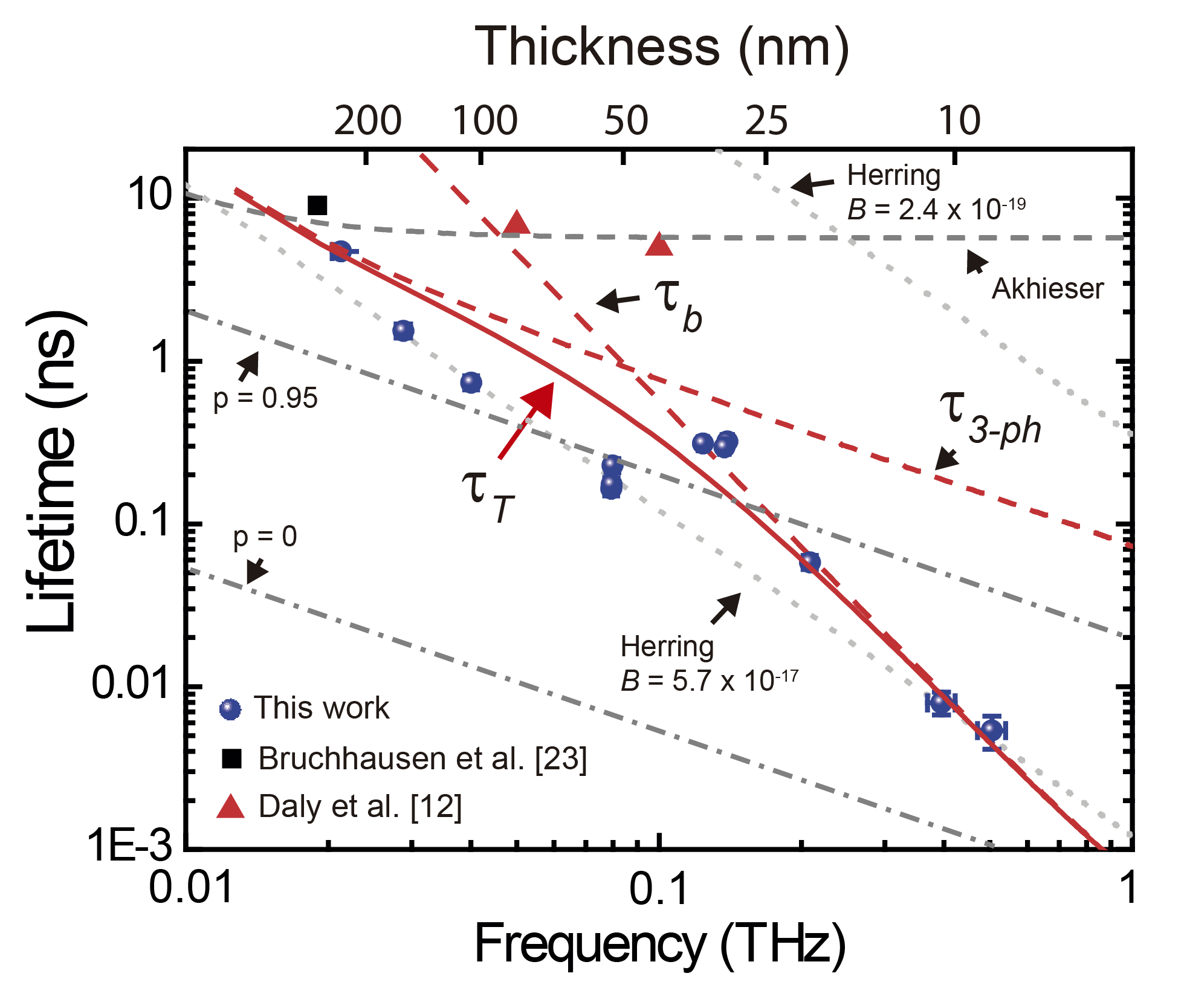}%
 \caption{\label{Fig3}Phonon lifetime of the first-order dilatational mode in free-standing silicon membranes as a function of frequency. Experimental data of free-standing silicon membranes with thickness values ranging from approximately 222 to 8 nm (black square \protect\cite{Bruchhausen2011}, blue circles) and bulk silicon (red triangles \protect\cite{Daly2009}). The red dashed lines show the contributions to the finite phonon lifetime from normal three-phonon interactions $\tau_{3-ph}$ and boundary scattering $\tau_{b}$ as indicated. The total contribution, calculated using Matthiessen’s rule $\tau^{-1}_{T}=\tau_{3-ph}^{-1}+\tau_{b}^{-1}$, is shown by the solid red line labelled $\tau_T$. Other models for intrinsic (grey dotted line: Herring  \protect\cite{Daly2009,Herring1954b}, grey dashed line: Akhiezer \protect\cite{Daly2009,Akhiezer1939}) and extrinsic (dot-dash grey lines: Casimir limit $p = 0$, $p=0.95$) scattering processes are shown for reference.}
 \end{figure}
 
In order to analyze our experimental data we first consider intrinsic damping mechanisms, which are inherent to even perfectly crystalline bulk materials. At high frequencies ($>$ 10 GHz) there are two main approaches to model the intrinsic phonon lifetimes due to the anharmonicity of the lattice. One commonly used model is that of Akhiezer relaxation damping, which considers the effect of the acoustic strain field on the populations of wavepackets of high-frequency phonons \cite{Akhiezer1939,Daly2009,Maris1971}. We found that this model as presented in Ref.\ \cite{Daly2009} does not reproduce the strong frequency dependence observed and overestimates the measured phonon lifetimes by at least one order of magnitude (Fig.\ 3, dashed grey line). The other commonly used approach to model intrinsic damping is a microscopic formulation considering three-phonon interactions, where the scattering probabilities are derived by applying first-order perturbation theory to a harmonic potential. The phonon-phonon scattering rates are generally derived under the single-mode relaxation time approximation, which assumes that during the decay of one phonon the other phonons maintain an equilibrium distribution, or equivalently, that the energy of the interacting phonon $\hbar \omega$ is large compared to the uncertainty in energies of the high-frequency phonons $\sim \hbar / \tau_{th}$ due to their finite lifetime $\tau_{th}$, i.e.\ $\omega _{th} \gg 1$ \cite{Maris1971}. Due to the great difficulty to evaluate  qualitatively the elements of the interaction matrix, early pioneering works \cite{Guthrie1966,Herring1954b,Holland1963,Klemens1958} made additional heuristic considerations regarding energy conservation surfaces and temperature regimes to arrive at convenient expressions of the frequency $\omega$ and temperature $T$ dependence of the phonon lifetimes. These expressions generally take the form:
\begin{equation}
\tau^{-1}=B T^{n} \omega^{m}
\end{equation}
where the parameters $B$, $n$ and $m$ are dependent on the temperature regime, polarizations of the interacting modes and crystal symmetry, and may be either approximated theoretically or empirically adjusted to fit experimental data. Results of this expression are shown in Fig.\ 3, with values $B_D = 2.4 \times 10^{-19} \text{ s } \text{K}^{-1}$, $n = 1$, and $m = 2$ as used by Daly et al.\ \cite{Daly2009} and Cahill et al.\ \cite{Cahill2005} for bulk silicon, derived from a fit to thermal conductivity data measured by thermo-reflectance. In Ref.\ \cite{Daly2009}, it was found that $B_D$ overestimated the measured bulk relaxation times. The best fit to our experimental data is obtained for a value of $B_{\text{\textit{EXP}}} = 5.7 \times 10^{-17} \text{ s } \text{K}^{-1}$, two orders of magnitude larger than $B_D$, which raises doubts about the validity of such an expression in our case. Moreover, the data clearly shows different trends in the studied frequency range. 

Here, we calculate explicitly the intrinsic scattering times under a Debye approximation, which we modify to consider specifically the $D^{1}_{q_{\parallel}=0}$ mode. Although the phonon cavity nature of the membrane causes a discretization of the out-of-plane acoustic spectrum \cite{Cuffe2012}, the Debye approximation neglects changes in the phonon density of states and therefore, this approximation can be expected to yield reasonable results for membranes thicker than $\sim$ 30 nm at room temperature \cite{Johnson1994,Lu2008a}. Notwithstanding the fact that this model does not include the effects of optical phonon modes, the dispersion of the bands for small wavelengths or acoustic anisotropy, it removes all adjustable parameters from the calculation with only the mode-averaged Grüneisen parameter not precisely known. As the $D^{1}_{q_{\parallel}=0}$ mode is purely longitudinal, we can express the relaxation time for a phonon with frequency undergoing normal three-phonon process of the type $\omega_L + \omega'_{s'} \rightarrow \omega''_{s''}$ as \cite{Srivastava1976}:
\begin{equation}
\begin{split}
\tau^{-1}_{3-ph}(\omega_L)&= \frac{\hbar v_L}{4 \pi \rho \bar{v}} \gamma^{2} \sum_{s',s''}\dfrac{1}{v^{2}_{s'}v^{2}_{s''}} \times \\ & \int \omega'_{s'}(\omega_L + \omega'_{s'})^{2}\frac{n(\omega'_{s'})(n(\omega''_{s''})+1)}{n(\omega_{L})+1)} d\omega
\end{split}
\end{equation} 

where $s$ is the mode polarization, $n$ is the Bose-Einstein distribution function, $\bar{v}$ is the phonon average group velocity and $\gamma$ is the mode-averaged Grüneisen parameter. In bulk silicon, the mode-dependent Grüneisen parameter lies in the range of 0.9 -- 1.3 \cite{Xu1991} for longitudinal modes. We take a value of 1.08 for the mode-averaged value, which has given the best fit  to thermal conductivity data of Si nanowires \cite{Omar2010}. Decay processes of the type $\omega_L \rightarrow \omega'_{s'} + \omega''_{s''}$ are not represented in Eq.\ (2) as they are unlikely to occur due to the low phonon energy and so have a negligible contribution to the total relaxation time [37]. The three-phonon interactions can be separated into those of the type $\omega_L + \omega'_{L} \rightarrow \omega''_{L}$ and $\omega_L + \omega'_{T} \rightarrow \omega''_{L}$, where $L$ and $T$ represent longitudinal and transverse polarizations, respectively. Collinear processes of the type $\omega_L + \omega'_{L} \rightarrow \omega''_{L}$ are sometimes neglected due to the dispersion of the branches, however, previous works have shown that these processes may occur as the finite lifetime of the branches compensates for the dispersion \cite{Maris1964a,Nava1964} and that they can play a large role especially at short wavevectors where the dispersion relation is quasi-linear. In fact, we find that $\omega_L \rightarrow \omega'_{L} + \omega''_{L}$ processes contribute most to the total intrinsic phonon lifetime.

The results of Eq.\ (2) are shown in Fig.\ 3 by the red dashed line labelled $\tau_{3-ph}$. We observe that this simplified theory yields the correct order of magnitude for phonon lifetimes in thicker membranes. However, the $\tau \propto \omega^{-1}$ frequency dependence is different from that exhibited by our experimental data for thinner membranes. While discrepancies at frequencies of $\sim$ 100 GHz and below could be related to uncertainties in the energies of the high-frequency phonons compared to the interacting phonon frequency, i.e., $\omega \tau_{th} \not\gg 1$, this relationship is expected to be well within its range of validity at higher frequencies. However, the experimental lifetimes are found to be orders of magnitude shorter than predicted. 

To explain this reduced lifetimes, we consider the impact of surface roughness scattering. We model its effect following the approach of Ziman \cite{Ziman1960}, where a single phenomenological parameter $p$ represents the ``polish'' of the surface, with $p=0$  for perfectly rough surfaces and $p=1$ for perfectly smooth surfaces. The $D^{1}_{q_{\parallel}=0}$ mode can be considered as a standing wave formed by the superposition of two counter-propagating longitudinal plane waves. We can then derive the wavelength-dependent specularity $p(\lambda) = \exp(-16 \pi^{3} \eta^{2} / \lambda^{2})$, where is $\eta$ the root mean square deviation of the height of the surface from the reference plane. Shorter wavelengths will therefore feel a stronger effect of the surface roughness than longer wavelengths. After considering a series of multiple reflections at the boundary, the mean free path can be written as $\Lambda = \dfrac{1+p}{1-p}\Lambda_0$ where $\Lambda_0$ is the characteristic dimension of the structure, i.e., the membrane thickness $d$ \cite{Ziman1960}. As a consequence, the lifetime due to boundary roughness scattering is
\begin{equation}
\begin{split}
\tau_b = \Lambda / v_L =& \frac{d}{v_L} \frac{1+\exp\left(-16 \pi^3 \eta^3/\lambda^2 \right)}{1-\exp\left(-16 \pi^3 \eta^3/\lambda^2 \right)} 
\\ =& \dfrac{d}{v_L} \coth \left(\frac{8 \pi^3 \eta^2}{\lambda^2} \right) \sim \frac{d}{\omega^2}\frac{v_L}{2 \pi \eta^2}
\end{split}
\end{equation}

The results of this expression are shown in Fig.\ 3 (red dashed line labelled $\tau_b$) and are compared to the frequently-used model considering a wavelength-independent specularity \cite{Srivastava1990}, with a value of $p = 0.95$ and to the Casimir limit of $p = 0$, shown by grey dot-dash lines. The experimental trend in lifetime as a function of frequency for the ultrathin ($>$ 30 nm) membranes is very well-described by the wavelength-dependent model  with a roughness value of $\eta$ = 0.5 nm. We observe that the phonon lifetime scales approximately as $\tau \propto \omega^{-3}$ due to the frequency-thickness relationship inherent to our sample set. We note that the native oxide layer on both sides of the membranes may introduce additional extrinsic scattering at the boundaries and we have included this effect empirically in the roughness value $\eta$. A simple combination of the lifetimes using Matthiessen rule $\tau_T^{-1} = \tau^{-1}_{3-ph}+\tau^{-1}_{b}$ appears to fit the lifetimes over the frequency range investigated.

By varying phonon populations and lifetimes, further temperature-dependent measurements should help to distinguish between the different scattering mechanisms. However, this work already provides much needed experimental data on phonon lifetimes in nanoscale systems at room temperature, for, e.g., direct use in the design of nano-mechanical oscillators and as input parameters for calculations of thermal conductivity in nanoelectronics and nanoscale thermoelectric materials. Furthermore, future work should also shed light on the predicted transition between Landau-Rumer and Akhiezer damping as the frequencies are in a suitable range \cite{Illisavskli1985,Maris1971}.

To conclude, we have shown experimental measurements of the relaxation times of coherent confined phonons in ultra-thin single-crystalline silicon membranes using the ASOPS technique, free from interference with a substrate or a deposited metal layer. The relaxation times of the ultra-thin membranes were found to be dominated by boundary roughness scattering which was modelled including a wavelength-dependent specularity. In the case of thicker membranes, phonon-phonon interactions were predicted to be the dominant scattering processes. The latter processes were calculated explicitly with a theory based upon three-phonon normal interactions, which gives the correct order of magnitude. However, further theoretical work is required to include the finite lifetimes of the high-frequency phonons in the three-phonon interaction model. We suggest that this may account for the discrepancies observed near 100 GHz and below where the phonon period becomes comparable to the lifetime of higher frequency phonons. Nevertheless, the preliminary combination of these theories seems able to predict phonon lifetimes in silicon membranes over several orders of magnitude up to 1 THz.

\begin{acknowledgments}
The authors acknowledge the financial support from the EU FP7 projects TAILPHOX (grant nr. 233883), NANOPOWER (grant nr. 256959), NANOPACK (grant nr. 216176) and NANOFUNCTION (grant nr. 257375); the Spanish MICINN projects ACPHIN (FIS2009-10150) and nanoTHERM (CSD2010-00044), the AGAUR 2009-SGR-150, and the Academy of Finland (grant nr. 252598). J.C. gratefully acknowledges a doctoral scholarship from the Irish Research Council for Science, Engineering and Technology (ICRSET) and E.C. gratefully acknowledges a Becas Chile 2010 CONICYT fellowship from the Chilean government.
\end{acknowledgments}

\bibliography{lifetimes}

\end{document}